# Formation of Monodispersed Cadmium Sulfide Particles by Aggregation of Nanosize Precursors

Sergiy Libert, Vyacheslav Gorshkov, Vladimir Privman, Dan Goia and Egon Matijević*

*Center for Advanced Materials Processing,*
*Clarkson University, Potsdam, NY 13699-5814, USA*

*Phone +1-315-268-2392, Electronic mail matiegon@clarkson.edu

**Abstract**

Monodispersed spherical cadmium sulfide particles were used as a model system in order to explain the size selection in the formation of colloids by aggregation of nanosize subunits. Several procedures of mixing the reactants were employed to precipitate these solids and follow the kinetics of particle growth. Efficient numerical simulation techniques for the model rate equations were developed to fit the experimental results. Our results have confirmed the recently proposed mechanism of two-stage growth by nucleation of nanosize crystalline primary particles and their subsequent aggregation into polycrystalline secondary colloids.

**Contents**

# 1. Introduction

The precipitation of submicrometer uniform spherical CdS particles was selected as an appropriate test system to explain the formation of monodispersed colloids by aggregation of nanosize primary precursors. Experiments were carried out under varying conditions to identify the optimal protocols and parameters of the processes involved, on the time scales that allowed the comparison with our recently developed theoretical model [1,2]. The latter was previously used to test the kinetics of the precipitation of uniform spherical gold particles. In the present work, we have explored extensions of the model, for a more complex system, CdS. A new numerical technique was developed to efficiently simulate the system of many rate equations encountered in modeling the particle growth.

While the preparation of many "monodispersed" inorganic colloids by precipitation from homogeneous solutions has been reported in the literature, the mechanisms of their formation have not been fully understood [3-6]. Early theoretical models assumed a short nucleation burst, followed by diffusional growth of the nuclei to form identical larger particles [7,8]. This mechanism works well for particles up to several tens of nanometers in diameter, though the resulting size distribution is not narrow. Many larger spherical particles, precipitated in homogeneous solution, showed polycrystalline X-ray characteristics, such as, ZnS [9], CdS [10], $Fe_2O_3$ [11]. Several techniques, including small angle light scattering, electron microscopy, and X-ray diffraction, have confirmed that these monodispersed colloids consisted of small crystalline subunits [9-19]. Furthermore, these subunits were of the same size [1,15,20] as the precursor particles of dimensions of order 10 nm, formed in solution, thus suggesting an aggregation-of-subunits mechanism. The above findings were not restricted to spherical colloids, but the same conclusion applied to particles of other shapes. Furthermore, it has been recognized that different morphologies of the final products must be related to the nature of the precursor subunits [11, 17-19].

The ultimate goal of a theoretical description should, therefore, explain the size-selection mechanism, i.e., the kinetics of the formation of particles of narrow size distribution, as well as the shape of the precipitated solids. Several approaches utilizing thermodynamic and dynamical growth mechanisms [1,5,6,17,21-32] have been proposed. Aggregation models were suggested that yielded a peaked particle size distribution [21-35], which may also sharpen with time. Here is further developed our new model [1], that accounts for the *size selection*, by coupling the rate equations for the processes of the final (secondary) particle formation by aggregation of the precursor subunits, to the dynamics of nucleation of these nanosize (primary) particles. The experimental system and techniques are described in Section 2. A new numerical approach to the model equations is developed in Section 3, while Section 4 presents the results and a summarizing discussion.

## 2. Experimental

In many instances, the formation of monodispersed colloids by precipitation involves nucleation and growth of nanosize primary precursor particles, which in turn aggregate into larger secondary colloid particles. The processes involved cover time scales from milliseconds to several hours. It is, therefore, crucial to investigate systems that are suitable for observing the time dependence and for controlling both of the dynamic processes involved. Precipitation of cadmium sulfide in aqueous solutions seems to offer definite advantages in these respects. The chemical composition of the final product is simple (CdS) and the solubility is low. Previously, it was shown [10] that under certain conditions, uniform spherical particles of desired size can be obtained by reactions at room temperature over a convenient time (several minutes). It was also documented that these spheres consisted of crystalline subunits. Furthermore, the procedure does not require any surfactants or other stabilizing agents. In principle, the process involves reacting cadmium salt solutions with thioacetamide, ($C_2H_5SN$, TAA), which decomposes in acidic environment and releases sulfide ions.

The sequence of chemical processes involved in the formation of CdS particles is presented below.

1) $Cd(NO_3)_2 \cdot 4H_2O \rightarrow Cd^{2+} + 2NO_3^- + 4H_2O$

2) $C_2H_5SN + HNO_3 \rightarrow S^{2-} + 2H^+ + \text{byproducts}$

3) $Cd^{2+} + S^{2-} \rightarrow CdS$

4) $mCdS \rightarrow (CdS)_m$

All chemicals were of the highest purity grade and used without further purification. Two different methods were employed in the precipitation of colloidal CdS particles: the controlled double jet precipitation (CDJP) and the seeding process.

In the CDJP, two reactant solutions are simultaneously introduced at a controlled rate, by means of peristaltic pumps, into a reactor prefilled with the desired solution, under continuous stirring. A number of reactant concentrations were tested in order to establish optimum conditions that would yield uniform spherical particles at reasonable times. Based on these preliminary results, the CDJP experiments were carried out with two reactant solutions prepared as follows:
The first solution was prepared in a 50 cm$^3$ volumetric flask, by dissolving 0.82 g of $Cd(NO_3)_2 \cdot 4H_2O$ in a small volume of double-distilled water, containing 0.99 g of nitric acid (70%). The flask was then filled to the mark with water, yielding a concentration of $5.32 \cdot 10^{-4}$ mol dm$^{-3}$ of $Cd^{2+}$, and 0.22 mol dm$^{-3}$ of nitric acid. The second solution was prepared in the same way, except that 0.2 g of TAA were dissolved, instead of the cadmium salt, which again corresponded to $5.32 \cdot 10^{-4}$ mol dm$^{-3}$ TAA and 0.22 mol dm$^{-3}$ HNO$_3$. All solutions were aged for about 15 hours at $T = 26°C$ before use.

In a typical CDJP experiment, 50 cm$^3$ each of the two reactant solutions, described above, were introduced into the reactor containing 100 cm$^3$ of the 0.22 mol dm$^{-3}$ HNO$_3$ solution at different pumping schemes, which are graphically illustrated in Fig. 1:

- In scheme # 1, the reacting solutions were rapidly mixed. They were delivered into the reactor within 1 second, then continuously stirred for 7 min. The CdS particles were then vacuum filtered and dried in a desiccator at 50°C for 5 h.
- In scheme # 2, the reactants were slowly added at the rate of 0.125 cm$^3$ sec$^{-1}$, for 400 sec. After 7 min, the solids were separated by filtration.
- In scheme # 3, the pumping rate was maintained constant at 0.55 cm$^3$ sec$^{-1}$. Thus, the solutions were pumped for 90 sec, but the separation of the solids was carried out after 7 min.
- In scheme # 4, the reactants were added in 10 cycles, each consisting of a 10 sec flow at 0.5 cm$^3$ sec$^{-1}$, with intervals of 20 sec. This process continued for 300 sec, and the separation was done after 7 min.

The second approach was based on earlier finding [10], that colloidal spherical CdS particles of very narrow size distribution could be obtained by a seeding procedure. In our experiments, seeds were obtained by a 16 h aging of a solution of 1.2·10$^{-3}$ mol dm$^{-3}$ of Cd(NO$_3$)$_2$, 0.3 mol dm$^{-3}$ of HNO$_3$, and 5·10$^{-3}$ mol dm$^{-3}$ of TAA, at 26°C. Then, 5 cm$^3$ of a 5·10$^{-2}$ mol dm$^{-3}$ TAA solution were added to 100 cm$^3$ of this seed dispersion. The resulting system was further kept for 100 min at 26°C.

The resulting CdS particles were characterized by scanning electron microscopy (SEM) and by powder X-ray diffraction. The roughness of the particle surfaces, which can be observed in the SEM images (Fig. 2) suggest that the particles are not monocrystalline. Figure 2a depicts the initial stage of the particle formation, taken after approximately 8 sec of precipitation according to Scheme #2, while Fig. 2b displays the same system after 1 min. To obtain these figures, the dispersion samples were withdrawn from the reactor with a microsyringe and rapidly filtered, a process that took less than 3 sec. The final product obtained by the same CDJP procedure, is shown in figure 2c. The particles surface appears quite smooth in contrast to those taken at earlier times. Finally, Fig. 2d shows particles obtained by the seeding technique.

The powder X-ray diffraction results for CdS particles prepared by different schemes described above, are presented on Fig. 3. The pattern is quite close to the one of natural mineral hawleyite [36], i.e., 3.36 (3.43), 2.068 (2.08), 1.75 (1.77), where numbers are the main crystal lattice spacings in Å for this mineral, and those in parentheses are for the synthesized particles. Using the Scherrer equation [37], the average subunit sizes were calculated to be 86 Å, 153 Å, 152 Å, 209 Å, for particles prepared by scheme #1, #2, #3, #4, respectively. These data confirm the crystalline nature of the nanosize subunits that form the spherical particles of CdS.

The histograms of the size distribution of particles prepared using different pumping schemes are presented in Fig. 4, and are based on digital sizing. Dotted lines are the fitted Gaussian distributions for these data, while solid lines are based on the theoretical model described below.

## 3. Theoretical

In synthesis of colloid particles, one aims at controlling their size distribution, as well as their shape and morphology. Until recently, there have been no successful models to explain these properties, and experimental parameters for achieving the uniformity of precipitated particles have been determined essentially by trial and error. Our new model [1,2] has explained semiquantitatively the size distribution of colloid gold particles [1,2,20]. Surprisingly, this was done without reference to particle shape and morphology. The key ingredients of the model, as well as the main assumptions, will be reviewed briefly below.

The primary particle formation is described by the nucleation model. The excess free energy of an embryo consisting of $n$ molecules of CdS, was modeled as the bulk term plus the surface term. The bulk term [1] results from the entropy of mixing, under the assumption that, owing to the covalent nature of the bonding, molecules, rather than individual atoms, are the dominant solute species in the embryo formation [38]. The surface term involves the effective surface tension of an $n$-molecule embryo. It is further assumed [1,2] that the solute transport is diffusional and that supercritical nuclei, are captured by the secondary particles without further contributing to the solute concentration balance. Then one can derive equations for the solute concentration, $c(t)$, and the rate of production of the primary particles, $\rho(t)$. These equations, detailed later, involve several parameters. Except for the surface tension, they can be all estimated from data available in the literature.

The secondary particle formation by aggregation was modeled [1,2] by rate equations that consider singlet (primary particle) attachment to dominate the process. Diffusional transport of the primary particles was assumed, with equations given below. Let us first qualitatively describe the mechanism of narrow size distribution in the secondary particle growth. After the initial batch of the secondary particles is formed, their growth is controlled by the supply of singlets. It turns out [1] that "protocols" of singlet supply can be realized which cause this batch of the secondary particles to grow in size without spreading significantly, while at the same time no longer generating many new secondary particles. Thus, the distribution does not really narrow, but its average size increases, and its *relative* width gets smaller.

The model involves several assumptions, which are discussed below. The main reason for these assumptions, some of which could be easily removed, has been to avoid too many adjustable parameters. For the primary particle formation and growth, the nucleation model ignores the fact that, especially in the early stages of the process, the critical nuclei consist of only few molecules. Therefore, the whole concept of the nucleation free energy, with the bulk plus surface terms, and the nucleation barrier, become questionable. However, the dynamics of such small clusters is not well understood [39]. Therefore, we use the nucleation model for all parameter values. Our model also ignores [1,2] the growth of the primary particles after they reach the critical nucleus size, but before they are consumed by the secondary particles. The reason for this has been that the primary particle radius only enters in the diffusional transport equations multiplied by the particle radius, in a combination that is not sensitive to the particle size. However, the matter balance is then violated, so our size distributions are only relative, and additional inaccuracies may result, that perhaps could be corrected by future more elaborate modeling.

The main assumption in the secondary particle growth has been that they irreversibly capture primary particles, and at the same time restructure to maintain compact, bulk-density, polycrystalline morphology. These assumptions, suggested by experimental observations, ignore many other possible dynamical processes, such as larger particle aggregates, detachment and ripening (singlet exchange), and direct consumption of molecules from the solution.

In the following subsections, we detail the model equations, developed and adopted for our experiments, following [1,2]. However, the main difficulty with using the coupled rate equations encountered in [1,2] has been the large computational effort required. Therefore, in this work, we have developed a new simulation approach that allows efficient evaluation of the secondary particle size distribution, and our presentation below is centered on the numerical method.

In our description of the model equation and parameters, we will focus on the numerical aspects of the evaluation of the secondary particle size distribution. Additional details of the model were reported in [1,2]. However, we also describe and substantiate extensions of the model, used here to improve consistency with the experimental data for CdS. In the original model, it was assumed that the number concentration, $N_k(t)$, of secondary particles, which consist of $k$ primary particles, evolves according to

$$\frac{dN_2}{dt} = N_1 (s_1 N_1 - s_2 N_2) ,$$

$$\frac{dN_{k>2}}{dt} = N_1 (s_{k-1} N_{k-1} - s_k N_k) ,$$

$$s_1 = 4 f \alpha ,$$

$$s_{k>1} = \alpha (1 + k^{1/3})(1 + k^{-1/3}) ,$$

(1)

where $N_1$ is the monomer (primary particle) concentration, the equation for which will be given later; $f$ is a correction parameter [2], originally equal $1/2$ to avoid double counting of monomers in Eq. (1), but eventually set to a smaller value for the physical reasons described later, in Section 4; $\alpha = 4\pi r D \zeta$ is the diffusional-capture-of-particles Smoluchowski rate expression, with $r$ and $D$ denoting the radius and diffusion coefficient of the primary particles; $\zeta \approx (0.58)^{-1/3} \approx 1.2$, where 0.58 is the typical filling factor of the random loose packing of spheres [2].

The process of capture of a primary particle proceeds its attachment at the surface of a secondary particle and then rapid incorporation in the structure, while still preserving its original crystalline inner core. The boundaries between the cores are filled, probably remaining amorphous, perhaps mediated by capture of molecular solute species, and the solvent is pushed out: the secondary particles experimentally have density comparable to the bulk material [3-6]. The shape and morphology of the secondary particles, including their surface morphology, such as the bumpy structures in Figs. 2a and 2b, which are much larger than the crystalline subunits, should be related to the dynamics of the primary particle incorporation. In many cases, smooth spherical particles are obtained, e.g., Figs. 2c and 2d. Here we take the expressions of the diffusional capture capacity of such spherical particles, consisting of a large number, $k$, of primary particles, and having radii $R \approx \zeta r k^{1/3}$. The diffusional flux of the primary particles to the surface is then proportional to the product $rD$ and, as mentioned earlier, is not sensitive to the radius of the primary particles, because $D \propto 1/r$.

Let us consider the continuum version of Eq. (1) for $k \gg 1$,

$$\frac{\partial N(k,t)}{\partial t} = N_1(t)\left[\frac{1}{2}\frac{\partial^2}{\partial k^2}(sN) - \frac{\partial(sN)}{\partial k}\right], \qquad (2)$$

where $N_1(t)$ is retained as a separate function of time. The "drift speed," $v$, of the distribution, along the $k$-axis, depends on the "coordinate," $k$, and time,

$$v(k,t) = N_1(t)s(k) = N_1(t)\alpha(1+k^{1/3})(1+k^{-1/3}) . \qquad (3)$$

We note that in the present model, only singlet (monomer) attachment processes were accounted for. Attachment of particles larger than primary, as well as detachment and monomer exchange processes (ripening) were neglected because they are slow and generally lead to broadening of the size distribution. However, Eq. (3) suggests that even monomer attachment somewhat broadens the distribution because $v$ is an increasing function of $k$. It has been emphasized [40] that it is the *relative* width of the secondary particle size distribution that can be made small in the two-stage growth process, rather than the absolute width.

A direct numerical solution of Eq. (1) is equivalent to one possible discrete approximation, on a grid with step $\Delta k = 1$, to the differential equation Eq. (2). This equation has a hidden numerical trap. Numerical solution of the system Eq. (1) by, for example, the Runge-Kutta technique, which uses the "explicit" recursion methods [41], is complicated by the fact that Eq. (2), which is basically a parabolic type equation with the "diffusion coefficient" $Q = N_1 s/2$, requires to satisfy stability condition [41], which in this case would limit the time step, $\Delta t$,

$$\Delta t < \frac{\Delta k^2}{2Q} = \frac{1}{N_1(t)s(k)} . \qquad (4)$$

Since $N_1(t)$ vanishes for large times, while the peak of the secondary particle size distribution freezes at fixed $k$, small time steps are required.

We have developed an implicit difference method [41], which has no such limitations on the choice of the time step in order to be stable. In addition, the number of nodes in the (time-dependent) spatial grid $k_{m=1,2,...}$ can be much smaller then the maximal number of primary particles in a secondary particle. It is possible to use an uneven grid, with the densest nodes in the regions where $N(k,t)$ varies significantly at time $t$. Our results were obtained by a variant of the Krank-Nicolson [41] method for parabolic equations, which is a second order approximation scheme in $\Delta k$ and $\Delta t$. This scheme is "conservative" in the sense defined in [41]. For an uneven spatial grid, we use the recursion relation

$$
\begin{aligned}
2\frac{N_m^{j+1}-N_m^j}{\Delta t} =& \frac{1}{\Delta_{av}}\left[\frac{Q_{m+1}N_{m+1}^{j+1}-Q_m N_m^{j+1}}{\Delta_m^+}-\frac{Q_m N_m^{j+1}-Q_{m-1}N_{m-1}^{j+1}}{\Delta_m^-}\right] \\
&-\frac{1}{\Delta_{av}}\left[v_{m+1/2}\frac{N_{m+1}^{j+1}+N_m^{j+1}}{2}-v_{m-1/2}\frac{N_m^{j+1}+N_{m-1}^{j+1}}{2}\right] \\
&+\frac{1}{\Delta_{av}}\left[\frac{Q_{m+1}N_{m+1}^j-Q_m N_m^j}{\Delta_m^+}-\frac{Q_m N_m^j-Q_{m-1}N_{m-1}^j}{\Delta_m^-}\right] \\
&-\frac{N_1}{\Delta_{av}}\left[v_{m+1/2}\frac{N_{m+1}^j+N_m^j}{2}-v_{m-1/2}\frac{N_m^j+N_{m-1}^j}{2}\right]
\end{aligned}
\qquad (5)
$$

where $\Delta_{av}=(\Delta_m^+ + \Delta_m^-)/2$, with $\Delta_m^+ = k_{m+1}-k_m$, $\Delta_m^- = k_m - k_{m-1}$; $k_m$ are the grid node coordinates; the upper index refers to the time slice, with $\Delta t = t^{j+1}-t^j$. The diffusion coefficient $Q(k_m)$ and drift velocity $v(k_m)$ depend on the concentration $N_1^{j+1/2} \equiv N_1(t^j + \Delta t/2)$, calculated separately, as described later. This scheme is reliable and easy to program. The values of $N(k_m)$ at each time slice are obtained from a system of linear equations, that form a tridiagonal matrix, and thus are easily solved by the technique described in [41].

Let us now consider the boundary condition for large $k$, namely, $N_{k\to\infty}(t\text{-fixed})=0$. In numerical simulations, this is realized by setting $N_{k\geq K(t)}=0$, where the $K(t)$ is calculated semi-empirically by the program as an estimate of the largest values of $k$ at time $t$ for which the distribution is nonzero. In order to optimize the calculation, it also possible to include in the program the procedure for periodic rebuilding of the difference grid for the variable $k$. The simulation process starts from the initial values $N_k(t=0)=0$.

The boundary conditions for Eq. (1,2) for small $k$, are given by the special form of the equation for $N_2(t)$, and by the equation for $N_1(t)$, which actually enters the equations explicitly and makes them nonlinear. The number of the primary particles, $N_1(t)$, is determined, in the continuum limit, by the equation

$$\frac{dN_1}{dt} = \rho(t) - 2s_{k=1}N_1^2(t) - N_1(t) \int_2^{K(t)} s(k)N(k,t)dk , \qquad (6)$$

where $\rho(t)$ is the rate at which the primary particles are nucleated, measured as number per unit volume per unit time. This relation simply represents the conservation of the primary particle number, counting those captured by the secondary particles, and follows from

$$\frac{dN_1(t)}{dt} = \rho(t) - \sum_{k=2}^{\infty} k \frac{dN_k(t)}{dt} , \qquad (7)$$

by using Eq. (1).

The rate of the primary particle generation by nucleation, $\rho(t)$, was calculated [1,2] based on the assumptions summarized in the Section 1. It depends on supersaturation level $\eta = c(t)/c_0$, where $c(t)$ is the number concentration of the molecular solute species, and $c_0$ is their equilibrium (saturation) concentration. The function $c(t)$ is the solution of the nonlinear differential equation

$$\frac{dc}{dt} = -\psi(c) + \xi(t) , \qquad (8)$$

where $\xi(t)$ is the rate of the solute species supply, which is determined by the experimental; recall the time-dependence protocols shown in Fig. 1. The rate of the concentration loss to nucleated primary particles, $\psi(c)$, is given [1] by

$$\psi(c) = \lambda c^2 (\ln \eta)^{-4} \exp[-\gamma(\ln \eta)^{-2}],$$

$$\lambda = 6.19 \cdot 10^4 a^9 \sigma^4 S(kT)^{-4}, \quad \gamma = 294 a^6 \sigma^3 S(kT)^{-3}, \qquad (9)$$

where $\sigma$ is the effective surface tension of the subcritical embryos, which can be usually set to the bulk value [1]. However, it is not known for CdS. Therefore, we have estimated it as an adjustable parameter, since the model results are sensitive to the choice of $\sigma$ [1], with the result $\sigma = (0.48 \pm 0.12)$ N/m, which is of a reasonable order of magnitude for minerals. We caution the reader that this value cannot be viewed as an unbiased estimate of the surface tension of CdS, because another parameter, $f$, was also adjusted, see below. There are two quantities in Eq. (9) that depend on the molecular dimensions of CdS, the diffusion constant of the solute species, $S$, and the effective radius, $a$, of the volume $\Omega$ per molecule in the crystalline structure of CdS, via $4\pi a^3/3 = \Omega$. Since earlier studies have indicated that the results are not sensitive to the values of these quantities, we took the rough estimate $a = 1.3 \cdot 10^{-10}$ m [38], and also used the same radius for the estimation of $S$ via the Stokes-Einstein relation.

The function $\rho(t)$ was evaluated in advance, before starting main part of the numerical calculation, via the relation [1]

$$\rho(t) = \psi(c) \left( \frac{8\pi a^2 \sigma}{3kT \ln \eta} \right)^{-3}, \qquad (10)$$

where the expression multiplying $\psi(c)$ is the number of molecules in the critical nucleus, which becomes a primary particle.

One can show that for a consistent second-order in time steps approximation, $N_1$ must be calculated in the intermediate nodes of the time grid. This was done by using

$$\frac{N_1^{j+1/2} - N_1^{j-1/2}}{\Delta t} = \rho^j - 2s_1 \left( \frac{N_1^{j+1/2} + N_1^{j-1/2}}{2} \right)^2 - \left( \frac{N_1^{j+1/2} + N_1^{j-1/2}}{2} \right) \int_2^K s(k) N^j(k) dk, \qquad (11)$$

where the notation for the indices $j, j \pm 1/2$ of the time grid was defined after Eq. (5). Eq. (11) is solved for $N_1^{j+1/2}$, and this value is used in order to advance from time slice $j$ to $(j+1)$; see Eq. (5).

As indicated by the experimental data, the number of the primary particles in a typical secondary particle can reach about $2 \cdot 10^6$. Our numerical procedure can evaluate the secondary particle size distribution with the $k$-grid of $3 \cdot 10^4$ nodes. The $k$-step was increased with $k$, in a geometric progression, with ratio $q = 1 + \delta$, where $\delta \approx 0.00024$. With time steps of order $10^{-3}$ s, the total calculation time for one parameter set did not exceed 1 hour on a 700 MHz PC.

Thus far, we have followed that model of [1] where the initial supply of the solute species was established before any nucleation or aggregation took place. In our case, there is an added complication that the total solution volume increases as a function of time owing to the addition of reagent solution according to the protocols shown in Fig. 1. Therefore, at each time step during which reactants are added, one should, in addition to iterating Eq. (5), recalculate all the concentrations according to the volume change. The latter was easy to determine, because the pumping rate, when nonzero, was always constant; see Fig. 1. Finally, for the lowest $k$ values, up to 100, the $k$-grid was exact, i.e., $k_m = m$.

## 4. Results and discussion

It has been noticed in earlier studies of the formation of monodispersed colloid gold, that the present model, if taken literally with the parameter $f = 1/2$ in Eq. (1), underestimates the secondary particle size. This implies that the model overestimates their number. The reasons for this can be several. Secondary particles, especially the smallest ones, can aggregate with each other. Another process not considered was ripening, which involves detachment of primary particles. Direct capture of solute matter can also play some role in the secondary particle growth. In [2], it was argued that instead of introducing numerous unknown parameters to account for all these effects, we can try to use the bottleneck factor to suppress particle production at the smallest size, $k = 2$.

In fact, there is another reason to set $f < 1/2$. Our definition of a primary particle refers to a monocrystalline entity. However, especially at early times, when the solute concentration is high, the critical nuclei are quite small. There is evidence that two such nuclei, when aggregating, can restructure into one monocrystal. Thus, the process will be schematically described by $1+1 \rightarrow 1$, instead of $1+1 \rightarrow 2$.

We found by numerical experiments that one actually has to assume $f \ll 1/2$. Since this possibility was not considered in earlier studies, we revisited the calculations for gold colloids [1,2]. The results, with the parameters for gold [1,2], are shown in Fig. 5, for $f = 5 \cdot 10^{-4}$, which significantly improves consistency with the experimental particle radius measured in [1], for their experimental time $t = 10$ s.

The same value, $f = 5 \cdot 10^{-4}$, also gave the results shown in Fig. 4 for CdS, with the parameters $a = 1.3 \cdot 10^{-10}$ m, $S = 4 \cdot 10^{-9}$ m$^2$s$^{-1}$, $\sigma = 0.48$ N/m, $c_0 = 8.78 \cdot 10^{16}$ m$^{-3}$. The particle size is predicted reasonably well. However, the calculated distributions are narrower than the experimentally measured ones, possibly indicating that the particle-particle aggregation and ripening processes should be really accounted for, beyond the introduction of the phenomenological parameter $f$, which we hope to address in future studies. Finally, Fig. 6 illustrates the time dependence of the secondary particle size distribution, for Scheme #1 of the reactant supply.

In summary, we have identified a system that allows experimental study of the formation mechanism of monodispersed colloids on reasonable time scales. For this CdS colloids, the present study has explored effects of the varying schemes of supply of reactants and compared the experimental results to the two-stage nucleation-aggregation model. We have also developed new numerical procedures to allow efficient numerical simulation of the model equations.

## Acknowledgements

This research has been supported by the National Science Foundation (grant DMR-0102644) and by the Donors of the Petroleum Research Fund, administered by the American Chemical Society (grant 37013-AC5,9).

**Figure Captions**

**Figure 1:** Pumping schemes used in the controlled double-jet precipitation (CDJP) of CdS. Scheme #1 corresponds to fast pumping at the rate of $50\,\text{cm}^3\,\text{s}^{-1}$, for 1 sec. Scheme #2 corresponds to slow pumping at the rate of $0.125\,\text{cm}^3\,\text{s}^{-1}$, for 400 sec. Scheme #3 is intermediate, with the rate of $0.55\,\text{cm}^3\,\text{s}^{-1}$, for 90 sec. Scheme #4 corresponds to stepwise pumping, which consists of 10 repeating cycles, with pumping at the rate of $0.50\,\text{cm}^3\,\text{s}^{-1}$ for 10 sec, and 20 sec interruptions.

**Figure 2:** Scanning electron micrographs (SEM) of CdS particles obtained by the controlled double-jet precipitation technique: **a**. after 8 sec; **b**. after 20 sec; and **c**. of the final product, all obtained according to Scheme #2; **d**. particles obtained by the seeding technique.

**Figure 3:** X-ray powder diffractions of CdS particles, obtained by the CDJP, employing different preparation schemes. A CuKα anode was used, with the X-ray wavelength $1.54\,\text{Å}$. The horizontal axis gives twice the scattering angle.

**Figure 4:** Experimental size distributions of colloidal CdS particles obtained by the CDJP process using different preparation schemes. The dashed lines are the calculated Gaussian fits, with the mean value and the standard deviation (SD) shown. The theoretical curves (solid lines) are based on the model calculations.

**Figure 5:** Evolution of the size distribution $N(k,t)$ in time, for gold colloid particles, calculated with the parameters given in [2], but with the same value, $f = 5 \cdot 10^{-4}$, as fitted here for CdS. The top panel gives the actual distribution, whereas the bottom panel gives the average particle radius as a function of time. For comparison, the experimentally measured average particle radius for gold was $1.0 \pm 0.1\,\mu\text{m}$ after 10 sec.

**Figure 6:** Example of the size distribution according to the secondary CdS particle radii, shown for times 0.1, 0.3, 0.5, 0.7, 0.9 and 20 seconds, from left to right, respectively, for Scheme #1.

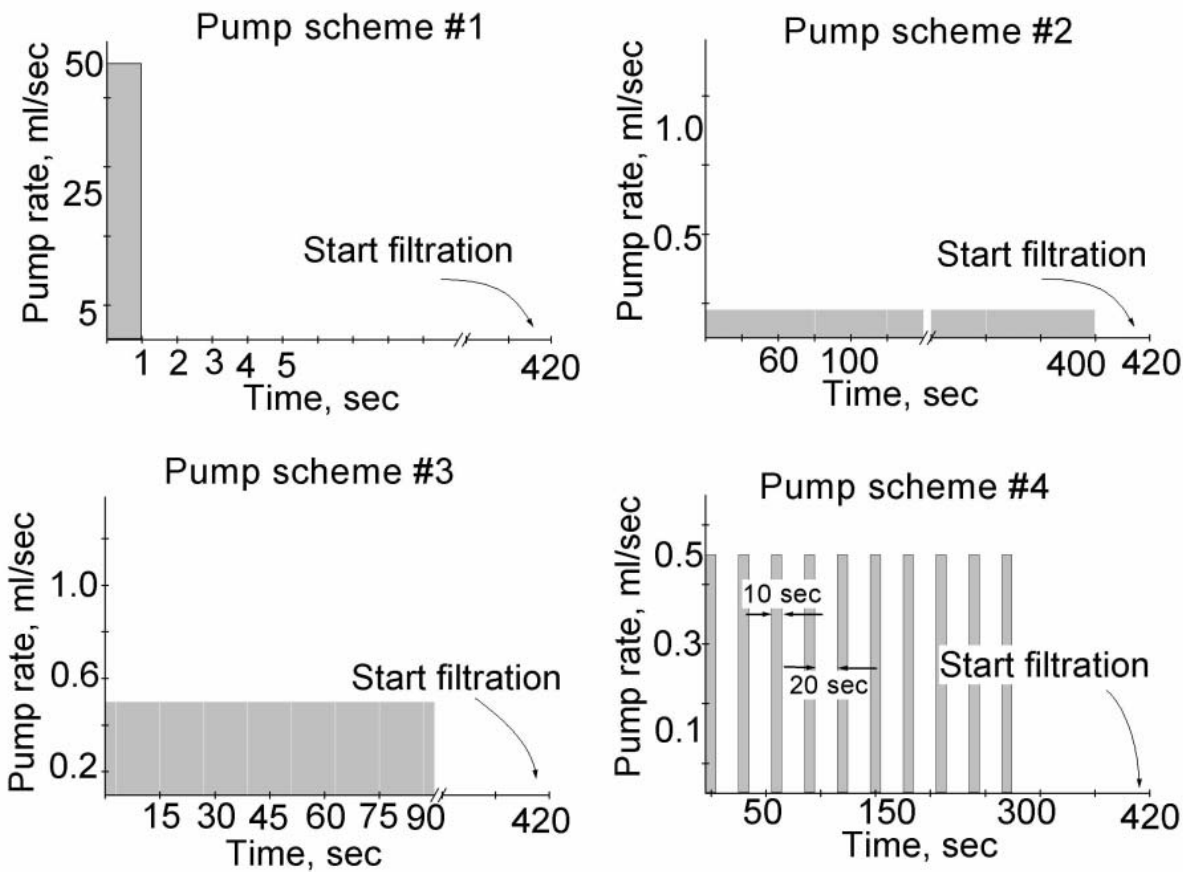

S. Libert, et al., Figure 1

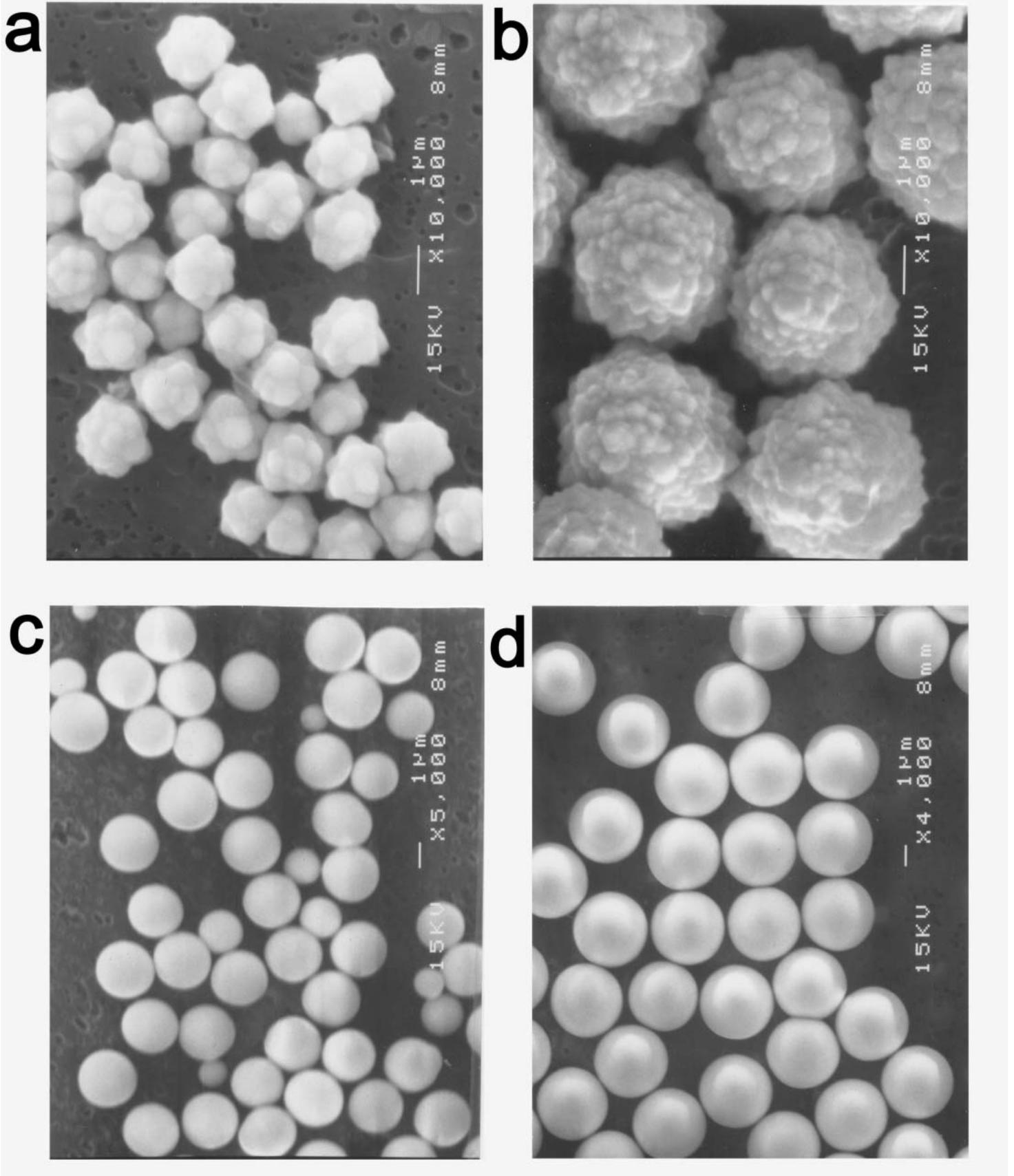

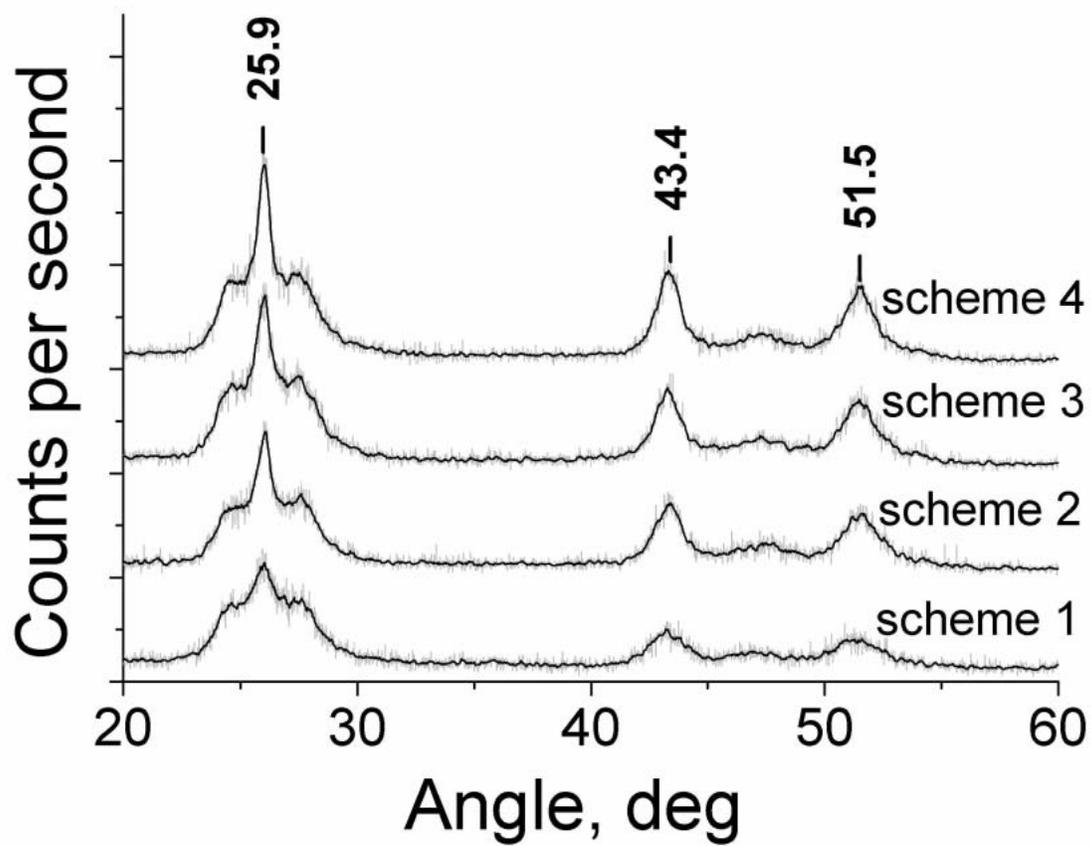

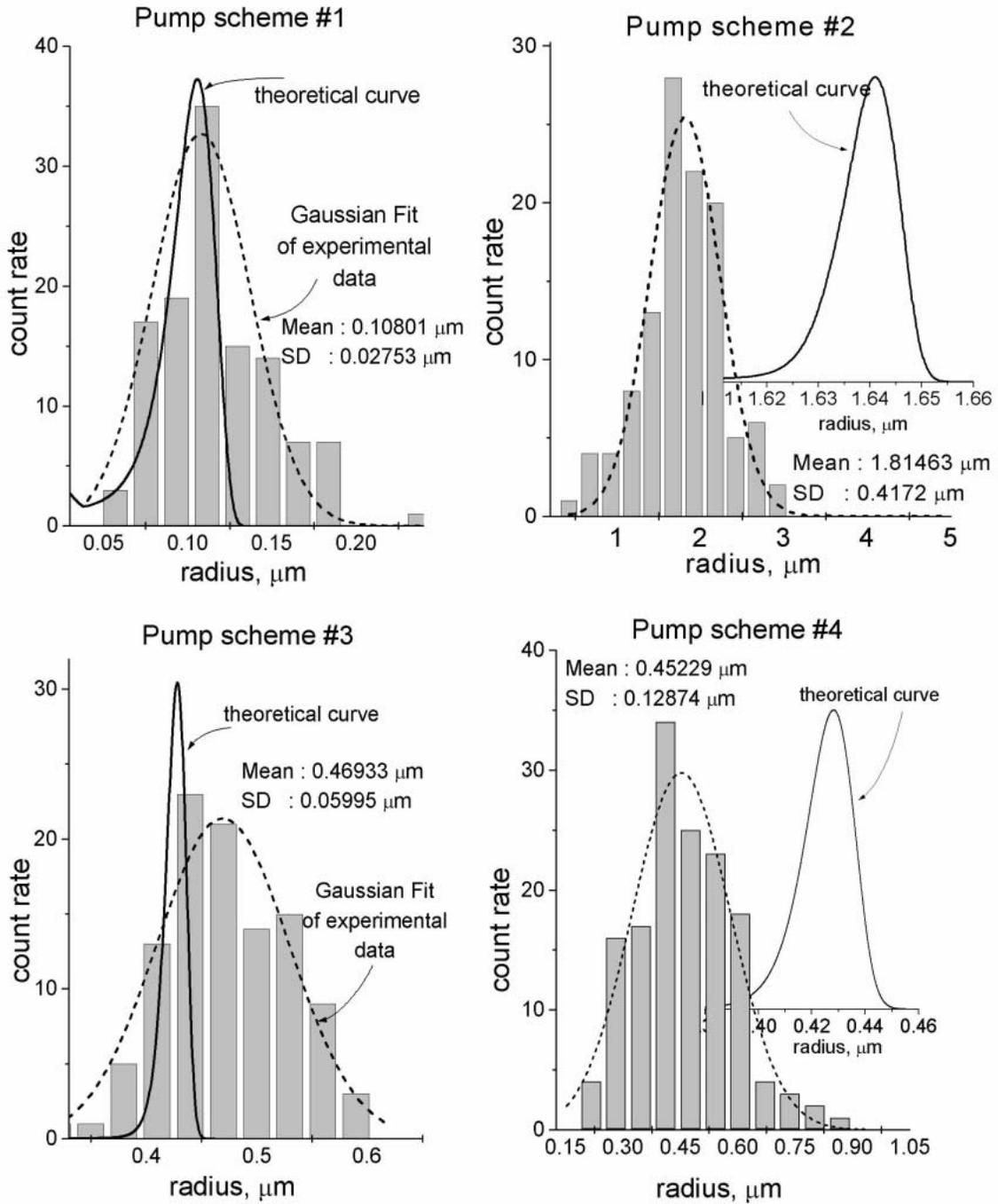

S. Libert, et al., Figure 5

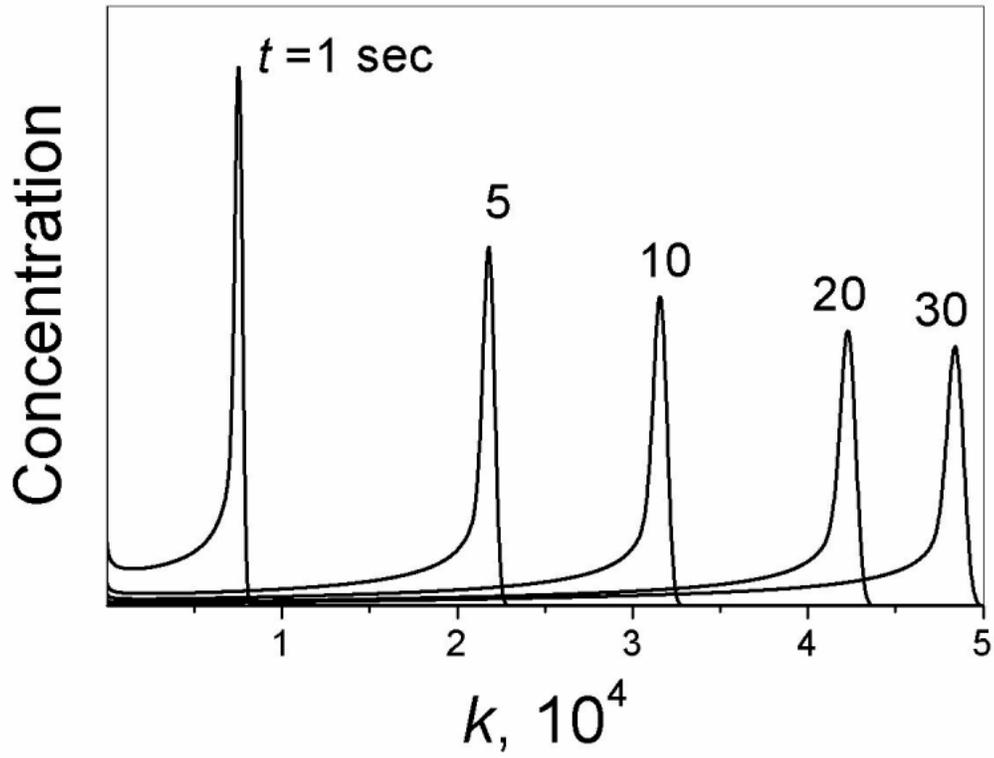
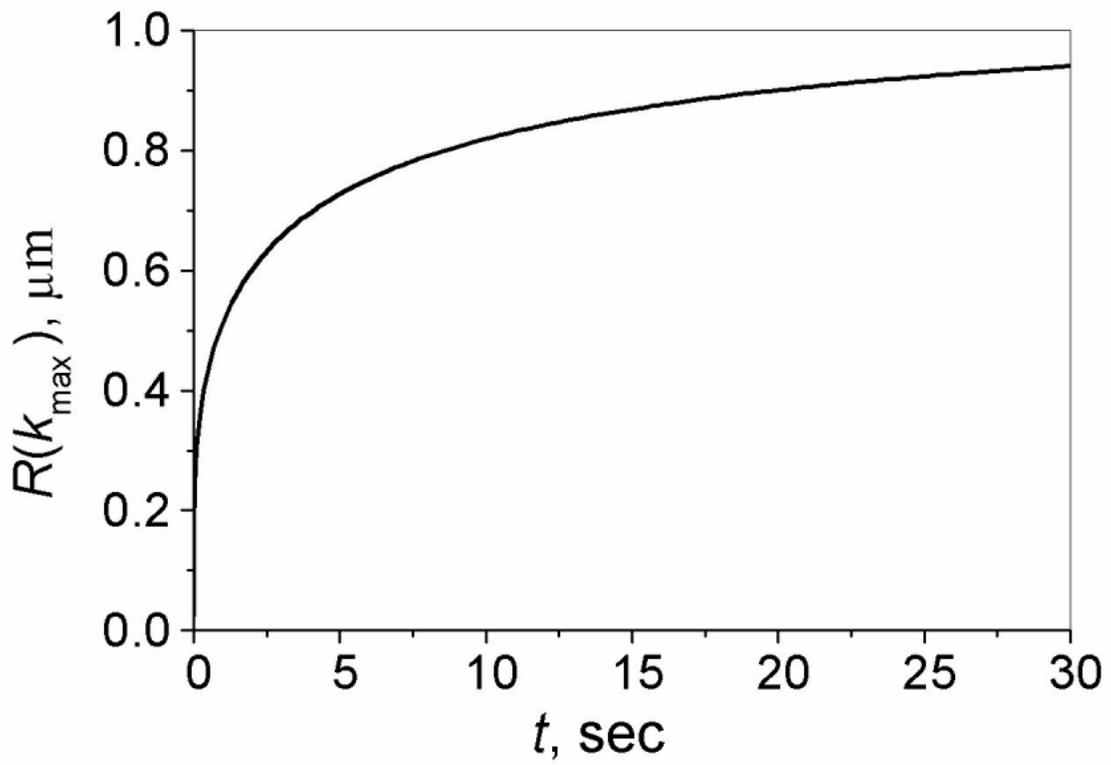

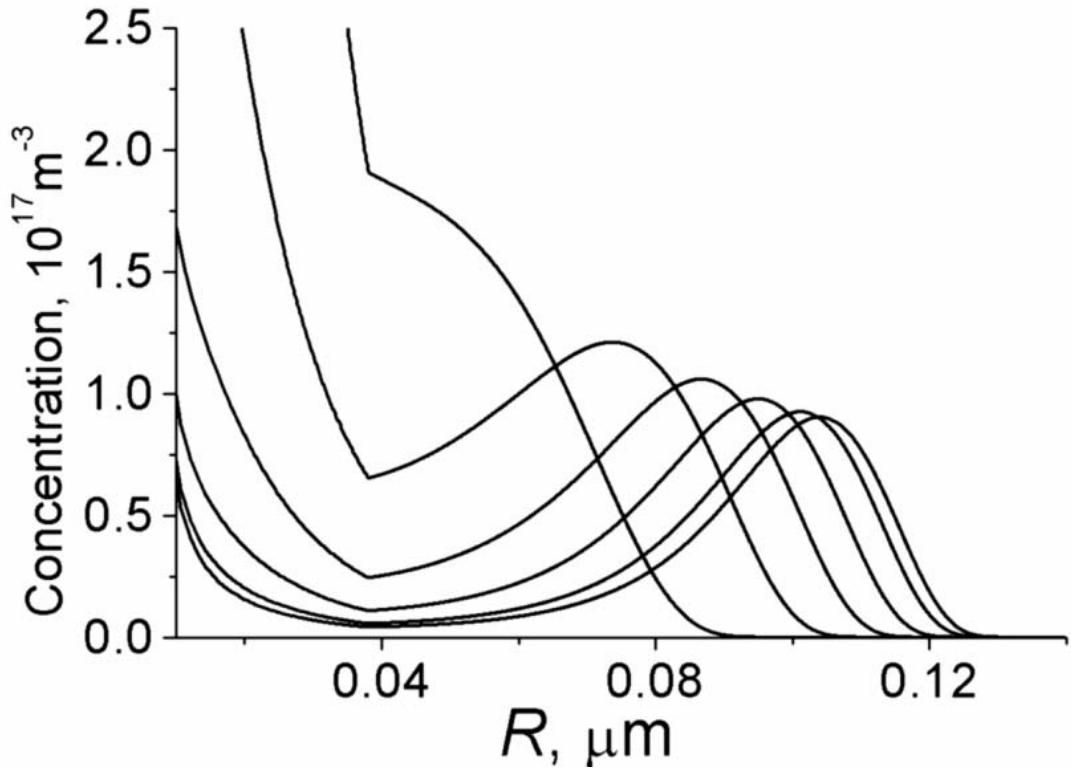

S. Libert, et al., Figure 6